# A Comparison Study of Coupled and Decoupled Uplink Heterogeneous Cellular Networks

Lan Zhang, Feng Gang, *IEEE, Senior Member*, Qin Shuang, *IEEE, Member*
National Key Lab of Communications
University of Electronic Science and Technology of China, Chengdu, China
Email：fenggang@uestc.edu.cn

*Abstract*— The evolution of mobile cellular networks has brought great changes of network architecture. For example, heterogeneous cellular network (HetNet) and Ultra dense network (UDN) have been proposed as promising techniques for 5G systems. Dense deployment of base stations (BSs) allows a mobile user to be able to access multiple BSs. Meanwhile the unbalance between UL and DL in HetNets, such as different received SINR threshold and traffic load, etc., becomes increasingly obvious. All these factors naturally inspire us to consider decoupling of uplink and downlink in radio access network. An interesting question is that whether the decoupled uplink (UL) /downlink (DL) access (DUDA) mode outperforms traditional coupled uplink (UL)/downlink (DL) access (CUDA) mode or not, and how big is the performance difference in terms of system rate, spectrum efficiency (SE) and energy efficiency (EE), etc. in HetNets. In this paper, we aim at thoroughly comparing the performance of the two modes based on stochastic geometry theory. In our analytical model, we take into account dynamic transmit power control in UL communication. Specifically, we employ fractional power control (FPC) to model a location-dependent channel state. Numerical results reveals that DUDA mode significantly outperforms CUDA mode in system rate, SE and EE in HetNets. In addition, DUDA mode improves load balance and potential fairness for both different type BSs and associated UEs.

## I. INTRODUCTION

Heterogeneous cellular networks (HetNets) have been recognized as one of the most promising approaches to meet the increasing requirements in 5G systems. When a cloud-like "data shower" appears in 5G, there will be increasing base stations (BSs). In the near future, the number of BSs may be comparable to the number of user equipment (UE) [1], where a UE may have multiple choice to associate with different BSs. Compared with traditional Macro-cell networks, HetNets introduce asymmetric between uplink (UL) and downlink (DL). For example the closest BS will achieve the maximum average received power in UL, but the maximum DL received power may come from far away macro BS [2]. Thus, the decoupled UL and DL access (DUDA) network structure is considered in 5G systems, where a UE can reasonably connect to a more appropriate BS, which may be different between UL and DL.

With the increase of symmetric traffic applications, such as video-calls and social networking, the improvement for quality of user experiences (QoE) in UL communication is well worthy studying. DUDA structure provides potential possibility to improve system performances. Meanwhile power control (PC) is a key consideration to improve UL system performances and reduce UE power consumption. 3GPP-LTE standard [3] has specified an optional closed-loop PC component around open-loop PC operation. The open-loop part, also named as fractional power control (FPC), is autonomously performed by UE to compensate long-term channel variation. The authors of [4] analyze PC in LTE UL by using simulations, and numerical results indicate that FPC can effectively improve throughput. The authors of [5] analyze the performance of open and closed-loop PC and examine the impact of system performance with maximum transmit power. The authors in [6] provide an analytical approach for FPC parameters selection.

In recent years, stochastic geometry has been a general tool for modeling wireless device distribution [7][8]. A number of researches apply stochastic geometry tools to analyze UL network performance of cellular networks. The authors in [9] propose a tractable model of uplink homogeneous cellular network with FPC consideration and evaluate the implications for power control. In [10], the authors evaluate EE in homogeneous system with dynamic TDD. In [11][12] a two-tier DUDA system is modeled for analyzing the system capacity and EE. A simple power adaptation model is applied without PC consideration. However, the decoupling model is dependent to the location of UE, which can be divided into 4 cases. Meanwhile the CUDA system model is not explicitly provided.

In this paper, we theoretically investigate the performance of DUDA and CUDA mode in HetNets. Using stochastic geometry tools, we construct a UL communication model of K-tier HetNet for DUDA and CUDA mode. We considers FPC scheme to model a location-dependent channel state with dynamic UE transmit power which is correlated with UE location. Numerical results reveals that compared with CUDA mode, the advantages of DUDA mode are threefold. First, DUDA mode brings load balance for UL communication in HetNets, which could be a severe problem especially in dense deployment CUDA mode. Second, DUDA mode significantly outperforms CUDA mode. In more detail, the system SE is improved by more than 100% and the improvement of EE is also very significant with FPC compensation. Third, DUDA mode provides fairness for UE associated with macro and small cells, and introduces stable system performance.

The rest of the paper is organized as follows. In Section II, we present our system model for CUDA and DUDA mode. In Section III, we derive the system rate, EE and SE of DUDA and CUDA modes in HetNets. In Section IV, we present numeral analysis of system performance in CUDA and DUDA system. Finally, we conclude this paper in Section V.

## II. SYSTEM MODEL

We consider a HetNet consisting of K-tier BSs denoted by $K=\{1,2,...,k\}$ arranged according to an independent homogeneous PPP $\Phi_{B_i}$ in the Euclidean plane. The differences of BSs across tiers include transmit power $P_{B_i}$, deployment intensity $\lambda_{B_i}$, and path loss factor $\alpha_i$. The standard path loss model is applied where $l(x) = \|x\|^{-\alpha_i}$ ($\alpha_i > 2$). UE locations are modeled by a different independent stationary point process $\Phi_U$ of intensity $\lambda_U$. Rayleigh fading is considered as *i.i.d.* random variables with exponential distribution $h \sim \exp(1)$. A simple case of uplink system model is illustrated in Fig. 1, where the UE in shadow area has different association in DUDA and CUDA mode. The dotted line represents coverage and connection in CUDA mode, and the solid line is for DUDA mode. The discrepancy of UL coverage areas for macro and femto BSs in CUDA mode is balanced in DUDA mode.

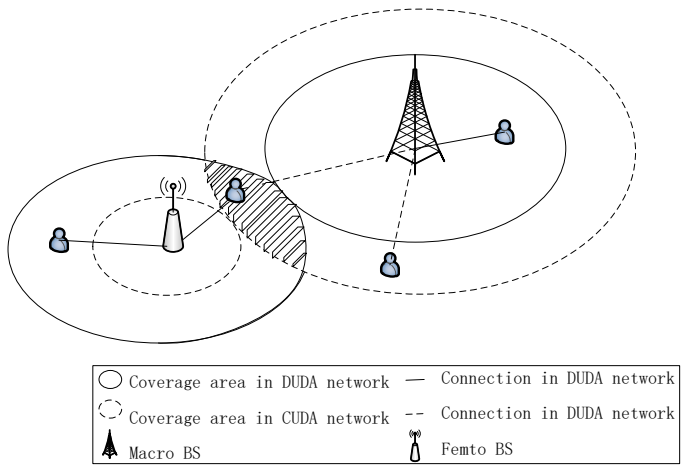

Fig. 1. Illustration of UL access scenario in DUDA and CUDA mode for a two-tier HetNet with a mix of macro and femto BS.

### A. System Assumptions

To facilitate our analysis, we use several assumptions as follows.

*1)* All BSs share the same total bandwidth, and the bandwidth is equally allocated among all UE.

*2)* All BSs are open accessed.

*3)* We assume UE power consumption model is separated into two fundamental parts $P_{UE} = P_{U_S} + P_{U_T}$, where $P_{U_S}$ is static power expenditure, and $P_{U_T}$ is transmitting power with upper bound $P_{U_T,\max}$.

### B. DUDA and CUDA HetNet Structure

In traditional CUDA structure, UEs are associated with the same BS with the strongest received power both in UL and DL. However, the UL-DL relationship in HetNets is quite different from homogeneous networks. A cell-edge UE may have poor UL SINR from the associated BS, but may have a better one in other BSs. Moreover different UL-DL interference may lead to different received SINR threshold, and the same access rule may not be appropriate. Also in UL, all UE are roughly equal power-constrained transmitters, and all BSs are just receivers. It is reasonable to design DUDA mode to figure out its potential advantages. In this paper we consider a type DUDA scheme, in which a UE is connected to the closest BS in UL. Other available decoupled access rules, such as connecting to BS with the minimum path loss, may be investigated in future work.

The distance from UE to its associated BS in the i*th* tier is denoted by a random variable $R_i$, $i \in K$. The probability distribution function (PDF) of $R_i$ is derived by null probability of a two dimensional PPP. $R_i^C$ represents $R_i$ in CUDA mode, and its PDF of is given by

$$f_{R_i^C}(r) = 2\pi\lambda_{B_i} r \exp(-\lambda_{B_i}\pi r^2), \quad i \in K \quad (1)$$

In DUDA mode, the UE is connected to the geographically closest BS. According to the superposition property of PPP, the independent k-tier PPP is equivalent to a homogeneous PPP $\Phi_{BH}$ of intensity $\lambda_{BH} = \sum_{i \in K} \lambda_{B_i}$. The PDF of $R_i$ in DUDA mode is $R_i^D$, whose PDF is given by

$$f_{R_i^D}(r) = 2\pi\lambda_{BH} r \exp(-\lambda_{BH}\pi r^2), \quad i \in K \quad (2)$$

### C. Power Consumption Model

According to assumption 3, we mainly analyze the effect of the dynamic transmit power. All UE are assumed to utilize distance-proportitional FPC to overcome path loss and reshape the distribution of interference power. The transmit power of the UE associated with the i*th* tier BS is given by

$$P_{U_T \sim B_i} = P_{U,\varepsilon_i} \cdot \|x\|^{\alpha_i \varepsilon_i}, \quad i \in K \quad (3)$$

where $P_{U,\varepsilon_i}$ is the baseline transmit power, whose value is constant and related with $\varepsilon_i$. When a UE is near to the desired BS, its transmit power can be decreased. $\varepsilon_i$ ($\varepsilon_i \in [0,1]$) is the FPC factor, which represents path loss compensation degree. Since the FPC factor is system design, we can adjust FPC factor to improve the whole system performances and provide better services for the majority of UE. When $\varepsilon_i = 0$ the transmit power is a static value $P_{U,0}$ without path loss compensation, and the FPC is full path loss compensation when $\varepsilon_i = 1$.

## D. SINR Model

According to Slivnyak's theorem [7], we analyse a randomly chosen BS located at origin (assumed in the i*th* tier). The distance between any UE and this BS is denoted by *D* with PDF given by

$$f_D(d) = 2\pi\lambda_U d \exp(-\lambda_U \pi d^2). \quad (4)$$

The received power of this BS from any associated UE is $P_{U,\varepsilon_i} h_u \|D\|^{\alpha_i(\varepsilon_i - 1)}$. Since the UE transmit power considers FPC scheme, we better model the location-dependent channel state. We consider that the UL interference come from a set of UE $v \in \Phi_I$, and the additive noise power is assumed as $\sigma^2$. The received SINR model of BS at origin is

$$SINR = \frac{P_{U,\varepsilon_i} h_u \|D_u\|^{\alpha_i(\varepsilon_i - 1)}}{I_I + \sigma^2}, \quad (5)$$

where

$$I_I = \sum_{v \in \Phi_I} P_{v,\varepsilon_v} h_v \|R_v\|^{\alpha_v \varepsilon_v} \|D_v\|^{-\alpha_v}. \quad (6)$$

## III. PERFORMANCE MODELING OF DUDA AND CUDA SYSTEM

In this section, we derive the EE and SE for UL of DUDA and CUDA networks. The EE and SE are respectively defined as

$$EE = \frac{\sum_{u \in \Phi_U} R_u}{\sum_{v \in \Phi_U} P_v}, \quad (7\text{-}1)$$

and

$$SE = \frac{\sum_{u \in \Phi_U} R_u}{W}, \quad (7\text{-}2)$$

where $R_u$ is the rate of UL transmission received by BSs, $P_v$ is the transmit power of UE. Since the transmit power can be adjusted by PC scheme, it is reasonable to consider EE of the transmit power. In the following we first analyze several essential factors for EE and SE analysis.

### A. Probability of Cell Association

The coupled and decoupled UL and DL access scheme directly affects the UL association scheme of UE. In DUDA mode, since a UE connects to geometrically closest BS, the association probability with a typical tier BS is only correlated with the intensity of that tier BS. Thus, we have Lemma 1 for cell association probability in DUDA mode.

**Lemma 1**: *The probability of UE association with the ith tier BS in DUDA network is given by*

$$A_i^D = \frac{\lambda_{B_i}}{\sum_{j \in K} \lambda_{B_j}} \quad (8)$$

In CUDA network, the UL association probability is much more complicated. Since a UE connect to the BS with the strongest received power, different tier of BS should be characterized by the spatial density, transmit power, bias path loss factor and fading. In [13] fading is averaged in cell association probability to achieve the long-term averaged received power. However in some fast-moving HetNet scenario, fading is still a general consideration. Thus we derive the cell association probability in two cases: considering fading effect.

**Lemma 2**: *The probability of UE association with the ith tier BS in CUDA network considering fading or not is respectively given by*

$$A_{i,case1}^C = \pi\lambda_{B_i} \int_0^\infty \int_0^\infty \frac{e^{-\pi \sum_{j=1, j\neq i}^k \lambda_{B_j} \left(\frac{P_{B_j}}{P_{B_i}}\right)^{\frac{2}{\alpha_j}} z^{\frac{2}{\alpha_j}} v^{\frac{\alpha_i}{\alpha_j}} - \lambda_{B_i} \pi v}}{(1+z)^2} dv dz \quad (9\text{-}1)$$

and

$$A_{i,case2}^C = \pi\lambda_{B_i} \int_0^\infty e^{-\pi \sum_{j=1, j\neq i}^k \lambda_{B_j} \left(\frac{P_{B_j}}{P_{B_i}}\right)^{\frac{2}{\alpha_j}} r^{\frac{\alpha_i}{\alpha_j}} - \lambda_{B_i} \pi r} dr \quad (9\text{-}2)$$

*Proof*: See Appendix A.

According to Lemma 1 and Lemma 2, we can derive the number of UE associated with the i*th* tier BS both in DUDA and CUDA system by $N_i = A_i \lambda_U$, and the number of the UE associated with one of BS in tier *i* by $n_i = A_i \lambda_U / \lambda_{B_i}$.

### B. Spatial Average Rate

According to assumption 1, bandwidth is equally shared by UE. The spatial average rate of UE associated with tier *i* is given by

$$R_i = E\left[\frac{W}{n_i} \ln(1 + SINR_i)\right].$$

**Theorem 1.** *The spatial average rate of UE associated with the ith tier BS in DUDA and CUDA system can be expressed by*

$$R_i = \frac{\pi W \lambda_{B_i}}{A_i} \int_0^\infty \int_{\ln(1+T_i)}^\infty L_I(s) \exp(-s\sigma^2) e^{-\lambda_U \pi r} dt dr \quad (10)$$

*where* $s = (e^t - 1)/(P_{U,\varepsilon_i} r^{\alpha_i(\varepsilon_i - 1)})$, $T_i$ *is the received SINR threshold in the ith tier BS, and* $L_I(s)$ *is the Laplace function of interference, which is different in DUDA and CUDA mode, and respectively given by*

$$L_{I_I}^D(s) = \prod_{j \in K} \pi\lambda_{BH} \int_0^\infty e^{-2\pi A_j^D \lambda_U \frac{Qr^{2-\alpha_j}}{\alpha_j - 2} {}_2F_1\left[1, 1-\frac{2}{\alpha_j}; 2-\frac{2}{\alpha_j}; -\frac{Q}{r^{\alpha_j}}\right]} e^{-\pi\lambda_{BH} u} du$$

*and*

$$L_{I_1^C}(s) = \prod_{j\in K} \pi\lambda_{B_j} \int_0^\infty e^{-2\pi A_j^C \lambda_U \frac{Qr^{2-\alpha_j}}{\alpha_j-2} {}_2F_1\left[1,1-\frac{2}{\alpha_j};2-\frac{2}{\alpha_j};-\frac{Q}{r^{\alpha_j}}\right]} e^{-\pi\lambda_{B_j} u} du,$$

where $Q \triangleq sP_{U,\varepsilon_j} u^{\frac{\alpha_j \varepsilon_j}{2}}$, and ${}_2F_1[\cdot]$ denotes the Gauss hypergeometric function.

*Proof*: The spatial average rate of the UE associated with the *i*th tier BS can be derived in a similar method in CUDA and DUDA mode. Here we only give the proof in DUDA scenario. When uplink SINR is smaller than the received threshold *T*, a transmission outage will occur. We take SINR threshold into consideration in (a), which is ignored in most correlated researches, where $t_0$ substitutes to the SINR threshold *T* to restrict the value of $ln(1+SINR)$.

$$R_i^D \stackrel{(a)}{=} \frac{W}{n_i^D} \int_0^\infty \int_{t_0}^\infty \Pr\left[\ln(1+SINR_i^D) > t\right] f_D(r) dt dr$$

$$\stackrel{(b)}{=} \frac{W\lambda_{B_i}}{A_i^D \lambda_U} \int_0^\infty \int_{\ln(1+T_i^D)}^\infty \Pr\left[SINR_i^D > e^t - 1\right] f_D(r) dt dr$$

Since the average rate is related with probability of SINR as shown in (b), we combine (5) and (6) and $h \sim \exp(1)$, we have

$$\Pr\left[SINR_i^D > e^t - 1\right] = \Pr\left(h_u > \frac{e^t - 1}{P_{U,\varepsilon_i} \|r^{(\varepsilon_i-1)}\|^{\alpha_i}} (I_1^D + \sigma^2)\right)$$

$$= L_{I_1^D}(s) \cdot \exp(-s\sigma^2)$$

To complete this proof, we then derive Laplace function of interference as below.

$$L_{I_1^D}(s) = E_{I_1^D}\left[e^{-sI_1^D}\right]$$

$$= E_{h,R,D}\left[\exp\left(-s \sum_{j\in K} \sum_{v\in(\Phi_1^D \sim B_j)} P_{v,\varepsilon_j} h_v \|R_v\|^{\alpha_v \varepsilon_v} \|D_v\|^{-\alpha_v}\right)\right]$$

$$\stackrel{(a)}{=} \prod_{j\in K} E_{R_j, D_j}\left[\prod_{v\in(\Phi_1^D \sim B_j)} \frac{1}{1 + sP_{v,\varepsilon_j} \cdot R_v^{\alpha_j \varepsilon_j} D_v^{-\alpha_j}}\right]$$

$$\stackrel{(b)}{=} \prod_{j\in K} E_{R_j}\left[\exp\left(-2\pi A_j^D \lambda_U \int_r^\infty \left(1 - \frac{1}{1 + sP_{U,\varepsilon_j} \cdot R_v^{\alpha_j \varepsilon_j} x^{-\alpha_j}}\right) x dx\right)\right]$$

$$= \prod_{j\in K} \pi\lambda_{BH} \int_0^\infty e^{-2\pi A_j^D \lambda_U \int_r^\infty \left(1 - \frac{1}{1 + sP_{U,\varepsilon_j} u^{\frac{\alpha_j \varepsilon_j}{2}} x^{-\alpha_j}}\right) x dx} e^{-\pi\lambda_{BH} u} du$$

$$\stackrel{(c)}{=} \prod_{j\in K} \pi\lambda_{BH} \int_0^\infty e^{-2\pi A_j^D \lambda_U \frac{Qr^{2-\alpha_j}}{\alpha_j-2} {}_2F_1\left[1,1-\frac{2}{\alpha_j};2-\frac{2}{\alpha_j};-\frac{Q}{r^{\alpha_j}}\right]} e^{-\pi\lambda_{BH} u} du$$

where (a) follows from the independence property of different fading channels and the exponential distributed fact, (b) follows from the probability generating functional (PFGL) of PPP [7]. We then substitute the inside integration of *x* into the hypergeometric function in (c), and finally obtain the desired result in (9).

The spatial average rate of the whole system in DUDA and CUDA mode can be derived based on Theorem 1, which is given by

$$R = \sum_{u\in\Phi_U} R_u = \sum_{i\in K} N_i R_i \quad (11)$$

### C. Spectrum Efficiency

According to definition in (7), we derive SE in DUDA and CUDA system in the following.

**Corollary 1**. *The average UL SE of UE belongs to the ith tier BS in DUDA and CUDA network is separately given by*

$$SE_i^D = 2\pi\lambda_{B_i}\lambda_U \int_0^\infty \int_{\ln(1+T_i^D)}^\infty L_{I_1^D}(s) e^{-s\sigma^2} r e^{-\lambda_U \pi r^2} dt dr \quad (12)$$

$$SE_i^C = 2\pi\lambda_{B_i}\lambda_U \int_0^\infty \int_{\ln(1+T_i^C)}^\infty L_{I_1^C}(s) e^{-s\sigma^2} r e^{-\lambda_U \pi r^2} dt dr \quad (13)$$

where $L_{I_1^D}(s)$ and $L_{I_1^C}(s)$ are Laplace function of interference and are the same as (10).

Corollary 1 can be easily proved with combination of (8-11). The whole system average SE in DUDA and CUDA network can be expressed as

$$SE = \frac{\sum_{u\in\Phi_U} R_u}{W} = \sum_{i\in K} SE_i \quad (14)$$

### D. Energy Efficiency

According to the definition of (7), we first analyze the average UE transmit power with FPC scheme to achieve EE expression.

**Lemma 3.** *In DUDA and CUDA network, the average transmit power of UE associated with the ith tier BS is separately given by*

$$E\left[P_{v,j}^D\right] = P_{U,\varepsilon_j} \left(\lambda_{BH}\pi\right)^{-\frac{\alpha_j \varepsilon_j}{2}} \cdot \Gamma\left(\frac{\alpha_j \varepsilon_j}{2} + 1\right) \quad (15)$$

$$E\left[P_{v,j}^C\right] = P_{U,\varepsilon_j} \left(\lambda_{B_j}\pi\right)^{-\frac{\alpha_j \varepsilon_j}{2}} \cdot \Gamma\left(\frac{\alpha_j \varepsilon_j}{2} + 1\right) \quad (16)$$

*Proof*: Since each UE applies FPC scheme to adjust their transmit power, the average UE power can be derived as $E\left[P_{U_T \sim B_j}\right] = \int_0^\infty P_{U,\varepsilon_j} r^{\alpha\varepsilon_j} f_{R_j}(r) dr$ based on the definition in (3). Combining with (1-2), we derive the average UE transmit power in (15) and (16). Combining Theorem 1 and Lemma 3, we can derive the expression of EE in DUDA and CUDA network.

**Corollary 2**. *The average EE of UE associated with the ith tier BS in DUDA and CUDA network is given by*

$$EE^i = \frac{\sum_{\substack{u \in \Phi_U \& \\ (v \text{ associate with} \\ \text{the } i\text{th tier BS})}} R_{u,i}}{\sum_{\substack{v \in \Phi_U \& \\ (v \text{ associate with} \\ \text{the } i\text{th tier BS})}} P_{v,i}} = \frac{R_i N_i}{E[P_{v,i}] N_i} = \frac{R_i}{E[P_{v,i}]} \quad (18)$$

*EE of the whole DUDA and CUDA network is given by*

$$EE = \frac{\sum_{i \in K} R_i \cdot N_i}{\sum_{j \in K} E[P_{v,j}] \cdot N_j} \quad (19)$$

## IV. NUMERICAL RESULTS

In this section, we numerically evaluate the performance of CUDA and DUDA mode of HetNet. We also present the analytical results to further validate our analysis. For clarity, our analysis is limited to an interference-limited two-tier HetNet. Other system parameters are given in TABLE II [14][15].

### A. Load Balance Brought by DUDA Mode

According to Lemma 1 and Lemma 2, we analyze DUDA and CUDA mode with the averaged fading association scenario (9-2). Based on the parameters in TABLE II, we illustrate association probability and average load ratio of macro cell and small cell in DUDA and CUDA mode in TABLE I. We consider two scenarios with the same macro BSs (MBS) and different types of small BSs (SBS) including femto and pico separately, where the femto scenario has more SBSs with lower transmit power.

TABLE I. COMPARISON OF DUDA AND CUDA NETWORK

| *Comparison Items* | *Pico scenario* | | *Femto scenario* | |
|---|---|---|---|---|
| | *DUDA* | *CUDA* | *DUDA* | *CUDA* |
| Association probability of MBS | 0.20 | 0.69 | 0.11 | 0.65 |
| Association probabality of SBS | 0.80 | 0.31 | 0.89 | 0.35 |
| Average load ratio(per MBS/per SBS) | 1:1 | 8.9:1 | 1:1 | 14.9:1 |

Since MBSs have over 20dBm transmit power more than SBSs, though the number of MBS is fewer, more UE is likely to associate with MBS in DL communication. Thus in CUDA system, although all BSs are equal receivers in UL communication, the UE must be associated to the same BS as DL communication. As illustration in TABLE I, the average UL load of a MBS is 7.9 times more than load of a SBS, and the gap is much bigger in femto scenario. DUDA mode solves this problem with uniformly distributed UE in macro and small cell. Thus the unbalanced UL load problem in CUDA mode disappears in DUDA mode.

### B. System Performance

As discussed above, DUDA mode in intuitions should outperform CUDA mode. In this part, we analyze system performances including SE and EE and examine the performance difference in the two modes. SE and EE value are correlated to several factors, including FPC factor $\varepsilon$, BS received SINR threshold and density of BSs and UE. We mainly consider macro and pico cell scenario with -10 dBm received SINR threshold to analyze SE and EE behaviors with the change of FPC factors. Refer to [4][5] that not all factors from 0 to 1 are applied in reality, and we mainly analyze FPC factors ranged 0.5-1 to evaluate SE and EE performance in DUDA and CUDA system.

TABLE II. PARAMETERS

| Parameters | value |
|---|---|
| Macro BS density ($/m^{-2}$) | 0.01 |
| Pico BS density ($/m^{-2}$) | 0.04 |
| Femto BS density ($/m^{-2}$) | 0.08 |
| UE density ($/m^{-2}$) | 0.20 |
| Macro BS path loss factor | 4.3 |
| Pico BS path loss factor | 3.8 |
| Femto BS path loss factor | 3.5 |
| Macro BS transmit power (dBm) | 43 |
| Pico BS transmit power (dBm) | 21 |
| Femto BS transmit power (dBm) | 17 |
| Max UE transmit power (dBm) | 23 |
| System bandwith (MHz) | 20 |

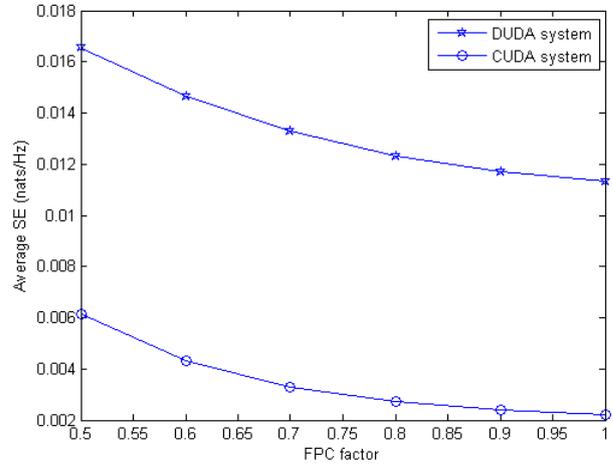

Fig. 2. Aversge SE of DUDA and CUDA system with the same FPC factor in macro and small cell

Fig. 2 shows the total network average SE with the same FPC factors of MBS and SBS. SE in DUDA mode is improved by more than 1 times. EE in these two modes have similar decreasing trend, which is mainly caused by lower transmit power of some UEs. Since the average total SE is composed by the UEs located in cell center and edge, though lager FPC factor limits interferences and increases rate for some UEs, the reduction of transmit power in other UE leads to decrease of total SE. Fig. 3 shows the total network average EE with the same FPC factors of MBS and SBS. EE in DUDA mode has a better performance. The improvement is obvious in small FPC factor value, and the difference decreases with more FPC compensation. FPC makes a tradeoff between mean cell throughput and cell edge throughput, the higher difference appears with lower compensation [4]. Thus the higher compensation the better balance throughput, which also decreases EE difference in DUDA and CUDA mode. In addition, different baseline power affects EE performance, which needs to be adjusted based on different circumstances. Thus the better behavior intuition of DUDA is confirmed.

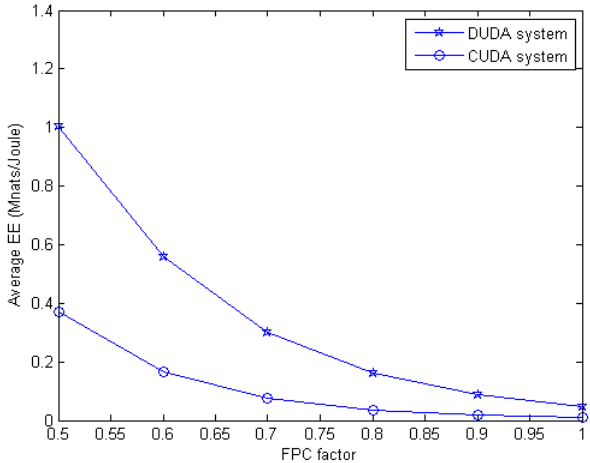

Fig. 3. Average EE of DUDA and CUDA system with the same FPC factor in macro and small cell

## C. Fairness for UE Associated with Different Cells

Note that the FPC factors of BSs in each tier are independently system-specific designed. Since part *A* figures out that DUDA mode brings UL load balance for macro and small cell, we are wondering whether a similar phenomenon will appear to UE associated with different type cells. In this part, we mainly analyze the potential goodness of DUDA mode brings to UE.

Fig. 4 shows the average rate of macro and pico cell with the same FPC factor in both MBSs and SBSs. Both two tier BSs in DUDA mode have a similar average rate, but the discrepancy appears obviously in CUDA mode. In CUDA mode, the average rate in macro cells is much lower than pico cells. The severe unbalanced cell load limits the average shared bandwidth of macro UE. Meanwhile, in CUDA mode the number of cell-edge UE is much more than DUDA mode, which will experience severe interference from intra and inter cell. Since in FPC scheme, cell-edge UE will have large transmit power, which will introduce much severer inter-cell interference. As shown in Fig.5, we examine the effect of average total network rate, when one FPC factor is static as 0.5, and the other in increased from 0.5 to 1. In DUDA mode, the effect of each tier FPC are relatively stable. While in CUDA mode, with the increase of MBS FPC factor, average total network rate increases nearly sevenfold. Since the majority UE are associated with MBS, the increase of transmit power improve the average total rate. However with the increase of the SBS FPC factor, the average total network rate is inversely decreased slowly. Although the increased SBS FPC factor increase the rate of UE associated with small cell, it introduces more interference which affects UE in macro cell especially the cell-edge UE. Since DUDA mode uniformly distributes UE for macro and small cell, the effect for cell-edge UE is decreased.

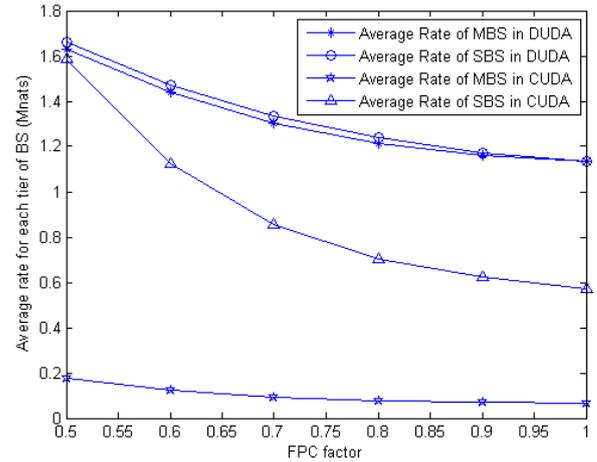

Fig. 4. The average rate in macro and pico cell in DUDA and CUDA system with the same FPC factor in different tier of BSs.

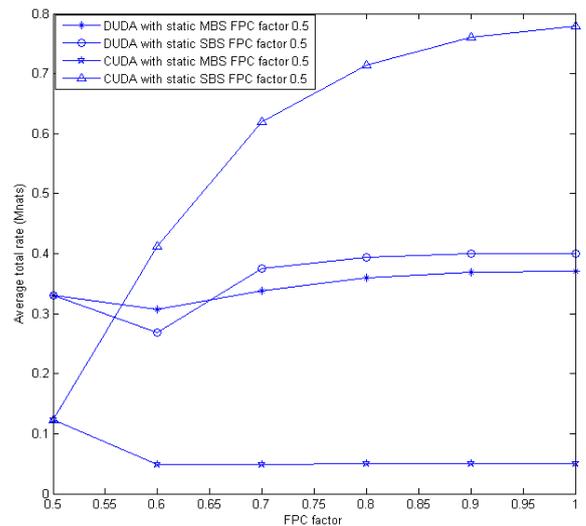

Fig. 5. The average total network rate in DUDA and CUDA system with a static FPC factor in one tier.

The results of Fig.4 and Fig.5 reveal that DUDA brings benefits not only for different type of BSs but also associated UEs. In both macro and small cell, all UEs use similar average rate, which is relatively stable. Also the average total network rate is not sensitive to the change of FPC factor in a type cell. DUDA mode brings fairness for UE associated with different type of BS, and introduce stable system performances.

## V. CONCLUSION

In this paper, we have conducted a system performances comparison study for decoupled and coupled UL/DL mode in HetNets. K-tier HetNet model of UL communication for these two modes are proposed using stochastic geometry tools. Our analysis applies FPC to provide dynamic per UE transmit power and model a location-dependent channel state. We mainly investigate system rate, SE and EE. Numerical results reveal that DUDA mode outperform traditional CUDA mode, which is not only the improvement of system SE and EE. Besides, DUDA mode brings load balance for HetNets, which is a severe problem of UL communication especially in dense deployment CUDA mode. Furthermore DUDA mode provides fairness for UE associated with macro and small cells to derive similar rate, and also leads to stable system performances.

## APPENDIX: PROOF OF LEMMA 2

We mainly give the proof for the general UE association probability with fading consideration. The proof of association probability without fading consideration is relatively easier, and can refer to [13]. According to Slivnyak's theorem, we analyze a typical UE located at origin associated with the ith tier BS.

$$A_{i\_f}^C = P\left[ P_{B_i}^C > \max_{j \in K, j \neq i} P_{B_j}^C \right]$$

$$= \prod_{j=1, j \neq i}^{k} E_{h_i, h_j, D_i} \left[ P\left[ P_i h_i \|R_i\|^{-\alpha_i} > P_j h_j \|R_j\|^{-\alpha_j} \right] \right]$$

$$\overset{(a)}{=} \prod_{j=1, j \neq i}^{k} E_Z \left[ \int_0^\infty P\left[ R_j > \left(\frac{P_j}{P_i}\right)^{\frac{1}{\alpha_j}} z^{\frac{1}{\alpha_j}} r^{\frac{\alpha_i}{\alpha_j}} \right] f_{R_i}(r) dr \right]$$

$$\overset{(b)}{=} \prod_{j=1, j \neq i}^{k} E_Z \left[ \int_0^\infty \exp\left[ -\lambda_{B_j} \pi \left(\frac{P_j}{P_i}\right)^{\frac{2}{\alpha_j}} z^{\frac{2}{\alpha_j}} r^{2\frac{\alpha_i}{\alpha_j}} \right] f_{R_i}(r) dr \right]$$

$$\overset{(c)}{=} \int_0^\infty \int_0^\infty e^{-\pi \sum_{j=1, j \neq i}^{k} \lambda_{B_j} \left(\frac{P_j}{P_i}\right)^{\frac{2}{\alpha_j}} z^{\frac{2}{\alpha_j}} r^{2\frac{\alpha_i}{\alpha_j}}} f_{R_i}(r) dr \cdot f_Z(z) dz$$

where (a) substitutes $h_i / h_j$ to z, and (b) follows from Poisson distribution of $R_j$. The PDF of Z is given by

$$f_Z(z) = f_{\frac{h_j}{h_i}}(z) = \begin{cases} \dfrac{1}{(1+z)^2} & z > 0 \\ 0 & z < 0 \end{cases}$$

then substitutes PDF of Z and $R_i$ in (c), and finally derives the desired expression in (9-1).